\documentclass{article}

\usepackage{graphics}

\begin{document}

\begin{center}
{\Large Resonance-enhanced group delay times in an asymmetric single quantum barrier}

\vspace{15pt} Chun-Fang Li$\footnote{Mailing address: Department of Physics, Shanghai University,
99 Shangda Road, Shanghai 200444, People's Republic of China. Email: cfli@staff.shu.edu.cn}$

\vspace{5pt}

Department of Physics, Shanghai University, 99 Shangda Road, Shanghai 200444, People's Republic of
China

State Key Laboratory of Transient Optics Technology, Xi'an Institute of Optics and Precision
Mechanics, Academia Sinica, 322 West Youyi Road, Xi'an 710068, People's Republic of China
\end{center}

\begin{abstract}
\normalsize{It is shown that the transmission and reflection group delay times in an asymmetric
single quantum barrier are greatly enhanced by the transmission resonance when the energy of
incident particles is larger than the height of the barrier. The resonant transmission group delay
is of the order of the quasibound state lifetime in the barrier region. The reflection group delay
can be either positive or negative, depending on the relative height of the potential energies on
the two sides of the barrier.  Its magnitude is much larger than the quasibound state lifetime.
These predictions have been observed in a microwave experiment by H. Spieker of Braunschweig
University.}

\vspace{10pt}

\noindent


\end{abstract}
\vspace{10pt}
\newpage

The tunnelling time of particles through single or multiple quantum barriers has drawn much
attention \cite{Hauge-S,Landauer-M,Nimtz-H, Chiao-S, Nimtz} with the advent of techniques for the
fabrication of semiconductor tunnelling devices, such as single-electron tunnelling transistors
\cite{Visscher-LVHMV}, resonant tunnelling diodes \cite{Brown-SPMMM}, quantum cascade lasers
\cite{Slivken-HER}, and resonant photodetectors \cite{Luna-HUGM}. Theoretical \cite{Buttiker-L,
Buttiker1, Ricco-A, Hauge-FF, Leavens-A, Steinberg-C, Paranjape, Japha-K, Winful, Buttiker-W} as
well as experimental \cite{Landauer-M, Landauer, Steinberg-KC, Nimtz-ES, Spielmann-SSF,
Ranfagni-MFP, Dekatskii-L, Dragoman-D, Martinez-P1} investigations have been extensively made on
this problem. It was found that the group delay (also referred to as the phase time in the
literature \cite{Hauge-S}) for particles tunnelling through a potential barrier saturates to a
constant value in the opaque limit \cite{Chiao-S, Balcou-D, Haibel-N, Martinez-P2}. This is the
so-called Hartman effect \cite{Hartman}. The reflection group delay is the same as the transmission
one in a symmetric configuration \cite{Steinberg-C}. It will be shown in this Letter that the
reflection group delay from a single barrier of asymmetric configuration can be negative and is
greatly enhanced by the transmission resonance.

The negative group delays in both reflection and transmission were previously discovered. But they
all occur in quantum-well structures, such as single quantum wells \cite{Li-W, Vetter-HN, Muga},
double-barrier quantum wells \cite{Grossel-DV, Buttiker2}, and their optical analogues \cite{Nimtz,
Longhi}. What we consider here is such a case in which particles are scattered by an asymmetric
single barrier, the height of which is less than the energy of incident particles. Quasibound
states were predicted and observed in such a situation \cite{Luo}. This is a classically allowed
motion. The particle in the barrier region has a real classical moving velocity which specifies a
classical traversal time $\tau_c$. It is found that the transmission and reflection group delays
are both enhanced by transmission resonance. The reflection group delay can be either positive or
negative, depending on the relative height of the potential energies on the two sides of the
barrier. The negative resonant reflection group delay corresponds to a transmission probability
that is larger than 1. The resonant transmission group delay is of the order of the quasibound
state lifetime in the barrier region and is larger than the classical time $\tau_c$. And the
magnitude of resonant reflection group delay is much larger than the lifetime of the quasibound
state in the barrier region. The reflected wave packet is considered without taking into account
the interference between the incident and the reflected waves \cite{Hauge-S, Grossel-DV}.

The height of the potential barrier, extending from $0$ to $a$, is $V_0$. The values of the
potential energies on the left and right handed sides of the barrier are $V_1$ and $V_2$,
respectively. It is assumed that $V_0>V_1$ and $V_0>V_2$. Let a beam of particles be incident from
the left, and let be $\psi_{in}(x)=A \exp(i k_1 x)$ the Fourier component of the incident wave
packet, where $k_1=[2 \mu (E-V_1)]^{1/2}/ \hbar$, and $\mu$ is the mass of incident particles. In
the following, we will assume that the energy, $E$, of incident particles is larger than the
height, $V_0$, of the potential barrier. Denoting, respectively, by $B \exp(-ik_1 x)$ and $F
\exp[ik_2 (x-a)]$ the corresponding Fourier components of the reflected and transmitted wave
packets, then the Schr\"{o}dinger equation and boundary conditions at $x=0$ and $x=a$ give $ r
\equiv \frac{B}{A}=\frac{g_1}{g_2} \exp[i(\phi_2-\phi_1)]$, and $t \equiv \frac{F}{A}=\frac{1}{g_2}
\exp(i \phi_2)$, where $k_2=[2 \mu (E-V_2)]^{1/2}/ \hbar$, $k_0=[2 \mu (E-V_0)]^{1/2}/ \hbar$,
non-negative number $g_1$ and real number $\phi_1$ are defined by a complex number as follows,
\begin{equation}
\label{phi1} g_1 \exp(-i \phi_1)=\frac{1}{2} (1-\frac{k_2}{k_1}) \cos (k_0 a)-\frac{i}{2}
(\frac{k_2}{k_0}-\frac{k_0}{k_1}) \sin (k_0 a),
\end{equation}
and non-negative number $g_2$ and real number $\phi_2$ are defined similarly by another complex
number as follows,
\begin{equation}
\label{phi2} g_2 \exp(-i \phi_2)=\frac{1}{2} (1+\frac{k_2}{k_1}) \cos (k_0 a)-\frac{i}{2}
(\frac{k_2}{k_0}+\frac{k_0}{k_1}) \sin (k_0 a).
\end{equation}
According to definition (\ref{phi1}), we have
\begin{equation} \label{tanphi1}
\tan \phi_1=\frac{1/k_0-k_0/k_1 k_2}{1/k_2-1/k_1} \tan (k_0 a),
\end{equation}
which shows that $\phi_1$ will change its sign by exchanging $k_1$ and $k_2$. This property will
have important effect on the group delay of reflected particles. According to Eq. (\ref{phi2}), we
have
\begin{equation} \label{tanphi2}
\tan \phi_2=\frac{1/k_0+k_0/k_1 k_2}{1/k_2+1/k_1} \tan (k_0 a),
\end{equation}
which shows that $\phi_2$ is symmetric between $k_1$ and $k_2$. We can also see from Eqs.
(\ref{tanphi1}) and (\ref{tanphi2}) that $\phi_1$ and $\phi_2$ can be exchanged from one to another
by changing the sign of $k_1$. This symmetry between $\phi_1$ and $\phi_2$ will simplify our
calculation of the group delay.

First, let us look at the group delay $\tau_t$ of transmitted particles. It is defined as $\hbar (d
\phi_2/ d E)$ \cite{Wigner, Steinberg-C} and is given by
$$
\tau_t=\frac{\tau_c}{4g_2^2} (1+\frac{k_2}{k_1})
[\frac{k_2}{k_0}+\frac{k_0}{k_1}-(1-\frac{k_0^2}{k_1^2}) (\frac{k_2}{k_0}-\frac{k_0}{k_2})
\frac{\sin 2 k_0 a}{2 k_0 a} ],
$$
where $\tau_c=a/v_c$ is the time taken for classical particles to travel through the barrier
region, $v_c=\frac{1}{\hbar (d k_0/d E)}=\frac{\hbar k_0}{\mu}$ is the classical velocity of the
particles in the barrier region. It is clear that $\tau_t$ is in general different from the
classical time $\tau_c$. Furthermore, it is easy to show that $\tau_t$ can be larger as well as
less than $\tau_c$. In fact, when $k_0 a=m \pi$ ($m=1, 2, 3\ldots$), $\tau_t$ reduces to
\begin{equation} \label{tautmax}
\tau_{tmax} \equiv \tau_t |_{k_0 a=m \pi}=\frac{k_1 k_2+k_0^2}{k_0 (k_1+k_2)} \tau_c.
\end{equation}
If the energy of incident particles is so close to the height of the potential barrier that $k_0$
is much less than $k_1$ and $k_2$, $\tau_{tmax}$ will be larger than $\tau_c$. On the other hand,
when $k_0 a=(m+1/2) \pi$, $\tau_t$ becomes $\tau_t |_{k_0 a=(m+1/2)
\pi}=\frac{1+k_2/k_1}{k_2/k_0+k_0/k_1} \tau_c$. It is less than $\tau_c$ under the above-mentioned
condition. In Fig. 1 is shown the dependence of $\tau_t$ upon the thickness $a$ of the barrier,
where $V_0/E=0.95$, $V_1/E=0$, $V_2/E=0.3$, and $a$ is re-scaled to be $k_0 a$. We see that
$\tau_t$ is maximum at $k_0 a=m \pi$.

\begin{figure}[ht]
\includegraphics{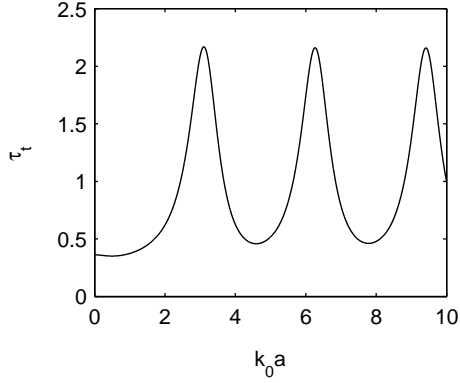}
\caption{The dependence of the transmission group delay $\tau_t$ upon the thickness $a$ of the
barrier, where $V_0/E=0.95$, $V_1/E=0$, $V_2/E=0.3$, $a$ is re-scaled to be $k_0 a$, and $\tau_t$
is in units of $\tau_c$.}
\end{figure}

It will be useful to introduce a quantity $T$, called transmission probability, which is defined as
\begin{equation}
T \equiv |t|^2=\frac{4 k_0^2 k_1^2}{k_0^2 (k_1+k_2)^2+(k_1^2-k_0^2) (k_2^2-k_0^2) \sin^2 k_0 a}.
\end{equation}
The transmission coefficient, defined as the ratio of the transmitted probability current density
to the incident probability current density \cite{Landau-L}, is then $(k_2/k_1)T$. When $k_0 a=m
\pi$, $T$ reaches its maximum, $T_{max}=\frac{4}{(1+k_2/k_1)^2}$. The resonance condition $k_0 a=m
\pi$ for transmission through a single barrier is the same as that for the quasilocalization of the
states in the barrier region \cite{Luo}. It can be shown that the resonant transmission time
(\ref{tautmax}) is of the order of the quasibound state lifetime in the barrier region
\cite{Ricco-A}. The symmetry of $\phi_2$ between $k_1$ and $k_2$ means that the transmission group
delay is the same whether the incident particles come to the barrier from left-handed or
right-handed side \cite{Steinberg-C}.

Now let us pay our attention to the reflection group delay, which is $\tau_r=\tau_t+\tau_1$ as is
seen from the expression of $r$, where
\begin{equation}
\tau_1=-\hbar \frac{d \phi_1}{d E}=-\frac{\tau_c}{4g_1^2} (1-\frac{k_2}{k_1})
[\frac{k_2}{k_0}-\frac{k_0}{k_1}-(1-\frac{k_0^2}{k_1^2}) (\frac{k_2}{k_0}-\frac{k_0}{k_2})
\frac{\sin 2 k_0 a}{2 k_0 a} ],
\end{equation}
and
\begin{equation} \label{g12}
g_1^2=\frac{1}{4}(1-\frac{k_2}{k_1})^2 \cos^{2} (k_0 a)
+\frac{1}{4}(\frac{k_2}{k_0}-\frac{k_0}{k_1})^2 \sin^{2} (k_0 a).
\end{equation}

Note:

(1) When the energy of incident particles is so close to the height of the potential barrier that
$k_1 \sim k_2$ and $k_0 \ll k_2$, the second term on the right-handed side of Eq. (\ref{g12}) is
usually much larger than the first term unless $k_0 a=m \pi$, at which $g_1^2$ is minimal. As a
result, near its maximum, $\tau_1$ can be approximated as
\begin{equation} \label{tau1approx}
\tau_1 \approx -\frac{\tau_c}{4g_1^2} (1-\frac{k_2}{k_1}) (\frac{k_2}{k_0}-\frac{k_0}{k_1}).
\end{equation}
And its maximum has a value of $\tau_{1max}=-\frac{k_1 k_2 -k_0^2}{k_0 (k_1-k_2)} \tau_c$.
Comparing with Eq. (\ref{tautmax}), we see that the magnitude of $\tau_{1max}$ is much larger than
the resonant transmission group delay, which means that the reflection group delay, $\tau_r$, is
dominated by $\tau_1$ near resonances, and its sign is determined by the sign of $\tau_1$. This
shows that the magnitude of the resonant reflection group delay is much larger than the quasibound
state lifetime in the barrier region. It is clear from Eq. (\ref{tau1approx}) that $\tau_1$ can be
negative as well as positive. When $k_1>k_2$ ($V_1<V_2$), $\tau_1$ (and hence $\tau_r$) is
negative. On the other hand, when $k_1<k_2$ ($V_1>V_2$), $\tau_1$ (and hence $\tau_r$) is positive.
These properties of $\tau_1$ can also be inferred from Eq. (\ref{tanphi1}). As a result, when the
reflection group delay for incidence from the left is positive, it is negative for incidence from
the right for the same configuration.

(2) When $k_1>k_2$, $T_{max}>1$. This shows that the negative peaks of the reflection group delay
correspond to a transmission probability that is larger than 1. The fact that the transmission
probability can be larger than 1 is not at odds with the law of probability conservation. It is the
probability current density, rather than the probability itself, that is in direct connection with
probability conservation in quantum scattering. In fact, the transmission coefficient,
$(k_2/k_1)T$, is always less than 1.

(3) The reflection coefficient $|r|^2$ does not vanish at the transmission resonance, so that the
reflected wave packet is well defined under the condition that follows (Eq. (\ref{restriction})).

(4) For the case of far from resonances, $k_0 a=(m+1/2) \pi$, $\tau_1$ becomes $\tau_1 |_{k_0
a=(m+1/2) \pi} = - \frac{1-k_2/k_1}{k_2/k_0-k_0/k_1} \tau_c$. Under the above mentioned conditions
($k_1 \sim k_2$ and $k_0 \ll k_2$), its magnitude is much smaller than the corresponding
transmission group delay, $\tau_t |_{k_0 a=(m+1/2) \pi}$. This shows that the reflection group
delay is almost the same as the transmission one when the energy of incident particles is far from
resonances.

In Fig. 2 is shown the dependence of $\tau_r$ upon the thickness $a$ of the barrier, where $a$ is
re-scaled to be $k_0 a$. The solid curve corresponds to negative-peak group delay, where all the
physical parameters are the same as in Fig. 1. The dashed curve corresponds to positive-peak group
delay, where $V_0/E=0.95$, $V_1/E=0.3$, and $V_2/E=0$. It is seen that the peaks of the group delay
occur at $k_0 a=m \pi$ and are much larger than the peaks of the transmission group delay whether
they are negative or positive.

\begin{figure}[ht]
\includegraphics{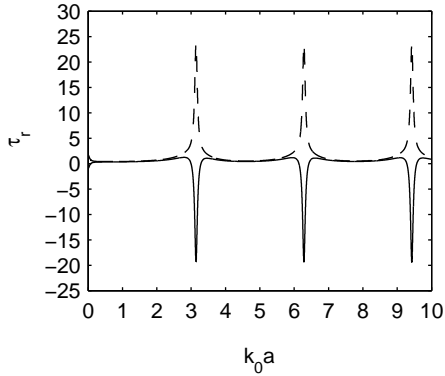}
\caption{The dependence of the reflection group delay $\tau_r$ upon the thickness $a$ of the
barrier, where $a$ is re-scaled to be $k_0 a$, and $\tau_r$ is in units of $\tau_c$. The solid
curve corresponds to the negative-peak group delay, where all the physical parameters are the same
as in Fig. 1. The dashed curve corresponds to the positive-peak group delay, where $V_0/E=0.95$,
$V_1/E=0.3$, and $V_2/E=0$.}
\end{figure}

The symmetries of $\phi_1$ and $\phi_2$ between $k_1$ and $k_2$ discussed before mean that the
transmission group delay and reflection group delay satisfy the average principle
\cite{Steinberg-C}, $\tau_{r1}+\tau_{r2}=2 \tau_t$, where $\tau_{r1}$ and $\tau_{r2}$ are
reflection group delays for the incident particles coming from left-handed side and right-handed
side, respectively.

Next, we discuss the validity of the above theoretical results. To this end, let us look at the
dependence of the reflection group delay on the energy of incident particles, which is shown in
Fig. 3, where $V_1=0$, $V_2=0.3 V_0$, $E \in [V_0,1.15 V_0]$, $a=10/(0.3 \mu V_0)^{1/2}$, and the
incidence energy $E$ is re-scaled to be $k_0 a$. The half width of the peak of the reflection group
delay, which can be approximately obtained from its dominant part $\tau_1$, is
$$
\Delta E=\frac{\hbar}{\tau_c} \sin^{-1} \frac{k_0 |k_1-k_2|}{[(k_1^2-k_0^2)(k_2^2-k_0^2)]^{1/2}}.
$$
For a Gaussian-shaped wave packet, its energy half-width $\delta E$ is related to its time
spreading $w$ by $\delta E \cdot w=\hbar/2$. For the above theoretical calculation to be valid,
that is, for the distortion of the reflected wave packet to be negligible, it is required that
$\delta E \leq \Delta E$. This results in a restriction on the thickness of the barrier,
\begin{equation} \label{restriction}
a \leq 2 v_c w \sin^{-1} \frac{k_0 |k_1-k_2|}{[(k_1^2-k_0^2)(k_2^2-k_0^2)]^{1/2}}.
\end{equation}

\begin{figure}[ht]
\includegraphics{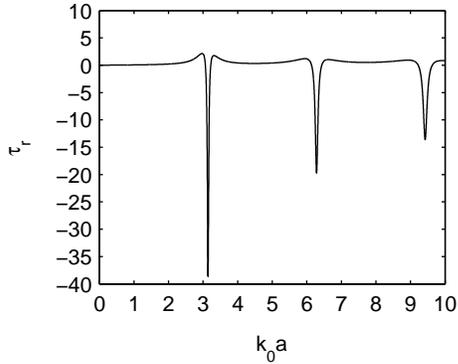}
\caption{The dependence of the reflection group delay $\tau_r$ on the energy $E$ of incident
particles, where $V_1=0$, $V_2=0.3 V_0$, $E \in [V_0,1.15 V_0]$, $a=10/(0.3 \mu V_0)^{1/2}$, $E$ is
re-scaled to be $k_0 a$, and $\tau_r$ is in units of $\tau_c$.}
\end{figure}

Because of the analogy between Schr\"{o}dinger's equation in quantum mechanics and Helmholtz's
equation in electromagnetism \cite{Chiao-S, Nimtz, Vetter-HN}, the predictions presented here have
been observed in a so-called G-band waveguide of width 47.5mm by H. Spieker of Braunschweig
University in Germany \cite{Spieker}, where the asymmetric barrier structure was obtained by
reducing the inside width of the waveguide, leading to effective widths of 40.5mm and 30.5mm. The
resonance-enhancement of the times is clearly shown, and both the positive and negative resonant
peaks of the reflection time is much larger than the resonant peak of the transmission time.

In a word, the reflection and transmission group delays in an asymmetric single quantum barrier are
greatly enhanced by the transmission resonance when the energy of incident particles is larger than
the height of the barrier. The reflection group delay can be negative as well as positive,
depending on the relative height of the potential energies on the two sides of the barrier. The
negative resonant reflection group delay corresponds to a transmission probability that is larger
than 1. The resonant transmission group delay is of the order of the quasibound state lifetime in
the barrier region and is larger than the classical traversal time. The magnitude of resonant
reflection group delay is much larger than the lifetime. These phenomena may have potential
applications in electronic devices, such as novel quantum-mechanical delay lines. It should be
pointed out that the negative reflection group delay does not imply a negative propagation
velocity. As a matter of fact, the negative group delay results from the reshaping \cite{Japha-K,
Dogariu-KCW} of the reflected wave packet due to the different phase shifts $\phi_2-\phi_1$ for its
different Fourier components.

\section*{Acknowledgments}
The author thanks G. Nimtz for his helpful discussions and suggestions. This work was supported in
part by the National Natural Science Foundation of China (Grants 60377025 and 60407007), Shanghai
Municipal Education Commission (Grants 01SG46 and 04AC99), Science and Technology Commission of
Shanghai Municipal (Grants 03QMH1405 and 04JC14036), and the Shanghai Leading Academic Discipline
Program.

\end{document}